\documentclass[12pt]{iopart}

\usepackage{graphicx}
\begin{document}

\title{Heavy Fermion Scaling: Uranium versus Cerium and Ytterbium Compounds}

\author{J. M. Lawrence$^1$, C. H. Wang$^2$, A. D. Christianson$^2$, E. D. Bauer$^3$}

\address{$^1$University of California, Irvine, CA 92697\\
  $^2$Oak Ridge National Laboratory, Oak Ridge, TN 37831\\
    $^3$Los Alamos National Laboratory, Los Alamos, NM 87545\\}

\ead{jmlawren@uci.edu}
\begin{abstract}
In an effort to explore the differences between rare-earth-based and uranium-based heavy Fermion (HF) compounds that reflect the
underlying difference between local 4$f$ moments and itinerant 5$f$ moments we analyze scaling laws that relate the low temperature
neutron spectra of the primary ("Kondo-esque") spin fluctuation to the specific heat and susceptibility. While the scaling appears
to work very well for the rare earth intermediate valence compounds, for a number of key uranium compounds the scaling laws fail
badly. There are two main reasons for this failure. First,  the presence of antiferromagnetic (AF) fluctuations, which contribute
significantly to the specific heat, alters the scaling ratios. Second, the scaling laws require knowledge of the high temperature
moment degeneracy, which is often undetermined for itinerant 5$f$ electrons.  By making plausible corrections for both effects,
better scaling ratios are obtained for some uranium compounds. We point out that while both the uranium HF compounds and the rare
earth intermediate valence (IV) compounds have spin fluctuation characteristic energies of order 5 - 25 meV, they differ in that
the AF fluctuations that are usually seen in the U compounds are never seen in the rare earth IV compounds. This suggests that the
5f itineracy increases the f-f exchange relative to the rare earth case.

\end{abstract}

%Uncomment for PACS numbers title message %\pacs{00.00, 20.00, 42.10} % Keywords required only for MST, PB, PMB, PM, JOA, JOB?
%\vspace{2pc} %\noindent{\it Keywords}: Article preparation, IOP journals % Uncomment for Submitted to journal title message
%\submitto{\JPA} % Comment out if separate title page not required

\maketitle

There are two categories of excitation seen in neutron scattering in paramagnetic heavy Fermion compounds. A "primary" spin
fluctuation gives rise, when excited, to the high temperature moment. Such an excitation, whose spectrum is typically a broadened
Lorentzian, is seen in all such compounds, from moderately heavy intermediate valence (IV) compounds to very heavy Fermion (VHF)
compounds. The latter compounds typically reside close to a quantum critical point (QCP) for a T = 0 transition to a magnetic
state, where magnetic fluctuations representing critical scattering close to the phase transition also occur.\cite{CFnote} In this
situation, scaling laws connect the behavior of the critical fluctuations to that of the specific heat and susceptibility. These
scaling laws have a different form for a QCP governed by "local" criticality as compared to the critical behavior of a magnetic
instability in a Fermi liquid.\cite{QCPcriticality}

One of the oldest-known properties of VHF and IV materials, whether in uranium or in rare earth (RE) compounds, is the existence of
a scaling law whereby the low temperature susceptibility $\chi (0)$ and specific heat coefficient $\gamma = C/T$ vary with the
inverse $1/T_{sf}$ of the characteristic energy $k_{B} T_{sf}$ of the primary spin fluctuation.\cite{Lawrence} The latter quantity
can be equated to the maximum $E_{max}$ in the dynamic susceptibility $\chi^{\prime\prime}(E)$, measured through inelastic neutron
scattering. An example is given in Fig. 1, where we compare the susceptibility, specific heat, and neutron spectra of two related
compounds, URu$_2$Zn$_{20}$ and UCo$_2$Zn$_{20}$.\cite{URu2Zn20} The zero temperature susceptibility and specific heat coefficient
of the latter compound are approximately three times larger, and the energy $E_{max}$ of the maximum in the neutron spectra is
three times smaller than in the former compound.

\begin{figure}[h]
\centering

\includegraphics[width=28pc]{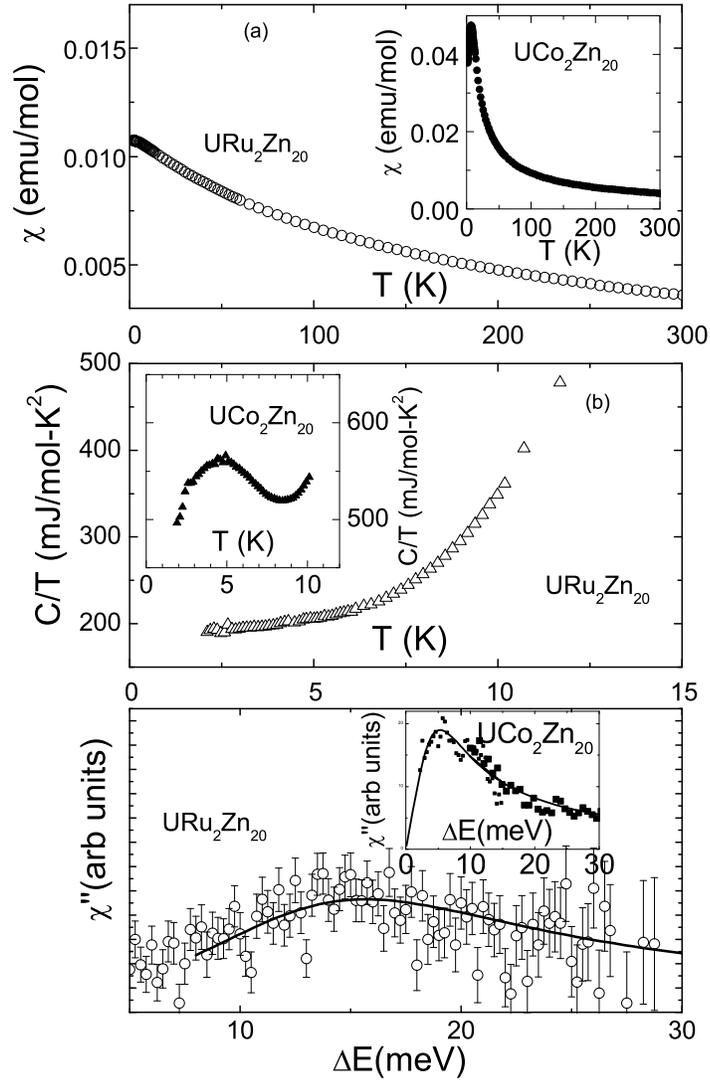}
\caption{\label{label} (a) Susceptibility, (b) specific heat coefficient and (c) inelastic magnetic neutron scattering spectrum of
URu$_2$Zn$_{20}$. The insets show the same quantities for UCo$_2$Zn$_{20}$. (Data from Ref. \cite{URu2Zn20})}

\end{figure}

The scaling laws arising from the primary spin fluctuation receive theoretical justification from the Kondo/Anderson impurity model
(K/AIM)\cite{Hewson}, where the spin fluctuation temperature $T_{sf}$ is identified with the Kondo temperature $T_K$. Despite the
fact that the rare earth atoms form a lattice, the K/AIM works very well to describe the susceptibility $\chi (T)$, specific heat
$C(T)/T$, 4$f$ occupation number $n_f$ (related to the valence through $z = 4-n_f$ for Ce and $z = 2 + n_f$ for Yb), as well as the
inelastic neutron lineshape for such IV compounds as YbAgCu$_4$\cite{SlowCrossover} and
YbFe$_2$Zn$_{20}$.\cite{URu2Zn20,CanfieldPNAS} A basic condition for this agreement between impurity theory and experiment is that
the 4$f$ spin fluctuations, which dominate these measurements, must be nearly localized, or only weakly dependent on momentum
transfer $Q$. This appears to be the case for such IV compounds as YbInCu$_4$\cite{LawrenceShapiro},
YbFe$_2$Zn$_{20}$\cite{URu2Zn20}, and CePd$_3$.\cite{FanelliPreprint} (Of course the K/AIM is inapplicable to measurements that are
highly sensitive to the lattice periodicity, such as electronic transport and de Haas van Alphen measurements.) Bevcause the K/AIM
theory works well in this context, we will refer to the primary spin fluctuation as "Kondo-esque."

In the VHF compounds the Kondo-esque fluctuation coexists with  magnetic (typically antiferromagnetic or AF) critical fluctuations
arising near the QCP, and both contribute to the specific heat and susceptibility. In what follows we will show that although the
Kondo-esque scaling initially appears to be invalid in uranium HF compounds, by correcting for the contribution of the AF
fluctuations, the  scaling behavior can be restored.

In angle-integrated photoemission studies rare earth 4$f$ states appear as localized states below the Fermi
level\cite{photoemission1}  but uranium 5$f$ states appear as a broad band of emission at the Fermi level.\cite{photoemission2}
Recent angle-resolved photoemission\cite{photoemission3} has determined the dispersion of the 5$f$ states. Essentially, the
4\emph{f} orbitals are highly localized and hybridize only weakly with the conduction electrons while the 5\emph{f} orbitals of
uranium compounds are spatially extended and form dispersive bands through strong hybridization with the neighboring $s$, $p$, and
$d$ orbitals.  One of our purposes is to see whether there are differences between the scaling behavior of 4\emph{f} and 5\emph{f}
heavy Fermion materials due to the distinction between local and itinerant $f$ electrons.

The Kondo/Anderson impurity model results in very precise scaling laws, but assumes localized electrons. Since we are concerned
with the applicability of scaling to itinerant 5$f$ electrons, we need to stress that the low temperature scaling laws follow from
very general considerations. The spin fluctuation peak at $E_{max} = k_{B} T_{sf} $ in the dynamic susceptibility represents an
excitation of the 4\emph{f} or 5\emph{f} moment out of a singlet (non-magnetic) ground state.  The excited level has the degeneracy
of the local 4\emph{f} (or itinerant 5\emph{f}) moment, which for rare earths is $N_J = 2J+1$. At high temperatures, the moment is
excited; at low temperatures, the moment "freezes out" and the system is nonmagnetic.

A phenomenology of the scaling can be obtained by making several simplifying approximations. The spin excitation at $E_{max}$ will
give rise to a peak in the specific heat at a temperature $T^C_{max}$ that is similar to a Schottky peak, with the exception that
$C(T)$ is linear at low temperature for heavy Fermion compounds. (This linearity is related to the substantial breadth of the
excitation.) If we assume that $C(T)$ is linear for $T < T^C_{max}$, that half the $R ln (2J+1)$ entropy is generated between $T =$
0 and $T^C_{max}$, and  that $T^C_{max}$ is approximately equal to $T_{sf}/$3 (a statement that is true for both Schottky anomalies
and for the Kondo specific heat), we then obtain $E_{max} \gamma =$ 3/2 $R ln (2J+1)$
 for the scaling constant. If we further assume that the susceptibility has a van Vleck-like form $\chi (0) \sim C_J / T_{sf}$
 (where $C_J$ is the free ion Curie constant for total angular momentum $J$) then we obtain the value 2 $\pi ^2/(9 ln(2J+1))$ for
 the Wilson ratio $W = (\pi ^2 R/(3 C_J))*\chi(0)/\gamma$.

It is clear that the scaling constants derived from this phenomenology can only be viewed as roughly approximate. The point of the
exercise is that scaling laws with values similar to those of the K/AIM (Table 1) are expected on very general grounds,
even when the spin fluctuations arise in an itinerant system where the K/AIM is not expected to be applicable. In particular, the phenomenology
exhibits the strong dependence of $E_{max} \gamma$ on the degeneracy. As we will see, this will allow us to distinguish
itinerant from local $f$ electrons.

 \begin{table}
\caption{\label{math-tab2}Comparison of scaling constants deduced from Kondo theory\cite{Rajan} and the rough phenomenology of the
text.}
\begin{tabular*}{\textwidth}{@{}l*{15}{@{\extracolsep{0pt plus
12pt}}l}} \br $J$&$E_{max} \gamma$(Kondo)&$E_{max} \gamma$(Rough)&$W$(Kondo)&$W$(Rough)\\ \mr

  &$(\pi /3) R J$ &(3/2) $R ln(2J+1)$&(1+(1/2$J$))&($2 \pi ^2 /9)/ln(2J+1)$\\

 3/2&   13.0&   17.3&   1.33&   1.58\\
 5/2&21.8&22.3&1.20&1.22\\
 7/2&30.4&25.9&1.14&1.05\\
 9/2&39.1&28.7&1.11&0.95\\

\br
\end{tabular*}
\end{table}

In Table 2, we exhibit the experimental values of $W$ and $E_{max} \gamma $ deduced from values of  $\chi(0)$, $\gamma$, and
$E_{max}$ obtained from the literature. In this table, we compare the uranium HF compounds to the rare earth IV compounds, for the
reason that the characteristic temperatures of the spin fluctuations are similar: they are typically 100 K or more, rather than
$\sim$ 10K as seen in the RE VHF compounds. We include all paramagnetic compounds in these two categories for which all three of the
quantities $\chi(0)$, $\gamma$, and $E_{max}$ have been reported. It can be seen that the low temperature scaling works well for
$J=$ 5/2 cerium and $J=$ 7/2 ytterbium IV compounds. Low temperature scaling also works well for URu$_2$Zn$_{20}$ and
UCo$_2$Zn$_{20}$, under the assumption that the high temperature moment is that of a Hund's Rule $5f^2$ or $5f^3$ electron. For
other classic uranium heavy Fermion compounds such as UAl$_2$, USn$_3$, UPt$_3$, and UBe$_{13}$ one or the other of the two scaling
constants is unacceptably different (a factor of two or more) from the value expected for the Hund's rule moment.

 \begin{table}
\caption{\label{math-tab2}Low temperature scaling constants for Ce ($J=$5/2), Yb ($J=$ 7/2), and U ($J=$ 4, 9/2 or undetermined)
compounds. Rows marked "Corrected" use the technique described in the text to correct for the AF fluctuation contribution to the
specific heat. (Sources of the data given in the first column.)}
\begin{tabular*}{\textwidth}{@{}l*{15}{@{\extracolsep{0pt plus
12pt}}l}} \br $Compound$&$\gamma$(J/mol-K$^2$)&$\chi (0)$ (emu/mol)&$E_{max}$(K)&$\gamma E_{max}$(J/mol-K)&Wilson ratio\\ \mr

  CePd$_3$\cite{KondoHole,VictorCePd3}&0.035&0.0018&638&22.3&1.74\\
CeSn$_3$\cite{LRP,MuraniCeSn3}&0.042&0.0018&464&19.5&1.49\\
\\
YbAl$_3$\cite{Cornelius,Christianson}&0.04&0.005&634&25.4&1.32\\
YbInCu$_4$\cite{SlowCrossover,LawrenceShapiro}&0.041&0.006&466&19.1&1.54\\
YbAgCu$_4$\cite{SlowCrossover}&0.199&0.017&133&26.5&0.9\\ YbFe$_2$Zn$_{20}$\cite{URu2Zn20,CanfieldPNAS}&0.52&0.05&70&36.4&1.02\\
\\
UAl$_2$\cite{BrodskyTrainor}&0.14&0.004&243&34.0&0.48\\ (Corrected)&0.07&&&17.0&1.30\\
USn$_3$\cite{Loong,Kambe}&0.17&0.01&60&10.2&0.99\\ URu$_2$Zn$_{20}$\cite{URu2Zn20}&0.188&0.012&191&35.9&1.08\\
UPt$_3$\cite{OttFisk,Goldman}&0.45&0.009&58&26.1&0.34\\ (Corrected)&0.225&&&13.0&1.10\\
UCo$_2$Zn$_{20}$\cite{URu2Zn20,BauerCo}&0.558&0.047&70&39.1&1.42\\ UBe$_{13}$\cite{OttFisk,LanderUBe13}&1.1&0.015&150&165&0.23\\
(Corrected)&0.183&&&27.4&1.38\\

\br
\end{tabular*}
\end{table}

The failure of the scaling laws for these uranium compounds arises from two problems. First, as discussed above, the low
temperature specific heat is affected not only by the excitation of the spin fluctuation at $k_B T_{sf}$ but by antiferromagnetic
(AF) fluctuations which occur when the compound resides close to a quantum critical point (QCP) for a $T =$ 0 transition between a
magnetic and a nonmagnetic state. Such AF correlations have been observed directly in neutron scattering experiments. These
fluctuations appear as highly $Q-$dependent peaks centered near the ordering wavevector \textbf{$Q_N$} of the antiferromagnetic
state; i.e., they are critical fluctuations. Typically, they occur on an order-of-magnitude lower energy scale than $k_B T_{sf}$
and are superimposed on a $Q-$independent (or weakly $Q-$dependent) background of scattering with energy scale $E_{max}= k_B
T_{sf}$.\cite{Goldman,Aeppli}  As an example of this behavior in U compounds, we show data for UPt$_3$ in Fig. 2. For the
scattering on the scale $E_{max}=$ 5-6 meV, the difference between scattering at zone center (0, 0, 2) and zone boundary (0, 0, 1)
can mostly be explained by the uranium 5$f$ form factor. This demonstrates that the primary spin fluctuation has a weak
$Q$-dependence which, as mentioned above, is known to be the case for RE IV compounds.   On the other hand, low energy  (0.3 - 0.5
meV) excitations peak sharply near the wavevector (1/2, 0, 1) of the weak antiferromagnetism that occurs in UPt$_3$.

 \begin{figure}[h]
\centering

\includegraphics[width=28pc]{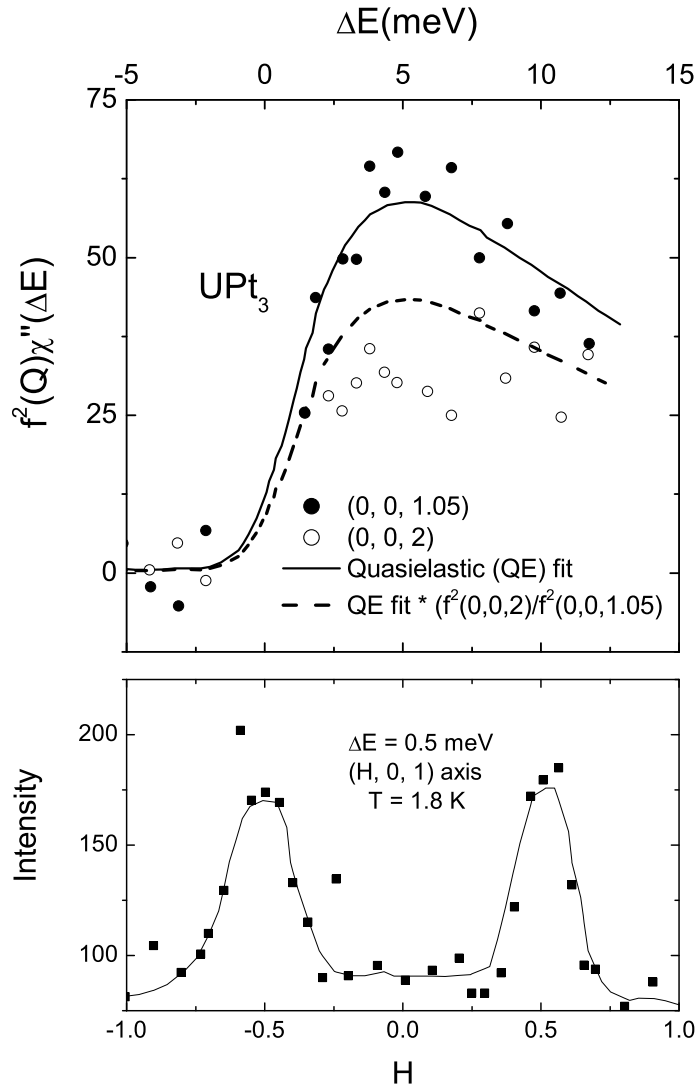}
\caption{\label{label} (a) Intensity versus energy transfer for UPt$_3$ at zone center (0, 0, 2) and close to the zone boundary (0,
0, 1.05). The solid line is a fit to a quasielastic power function with $\Gamma = $ 5 meV; the dashed line is the same fit scaled
by the ratio of the 5$f$ form factor for the two $Q$. (b) Intensity versus Miller index H along the [H, 0 1] direction for energy
transfer 0.5 meV. The peaks at (1/2, 0, 1) represent antiferromagnetic fluctuations. (Adapted from Refs. \cite{Goldman, Aeppli})}

\end{figure}

These AF fluctuations can give rise to a low temperature upturn in the specific heat coefficient $C(T)/T$ on the same temperature
scale, which enhances the specific heat above the scaling value.  The scaling in uranium compounds can be corrected by subtracting
this contribution, or rather by extrapolating the specific heat coefficient from temperatures above the upturn. The extrapolation
is shown in Fig. 3 for UBe$_{13}$.  The low temperature specific heat coefficient $\gamma =$ 1.1 J/mol-K$^2$ gives $E_{max} \gamma
=$ 165 J/mol-K and $W$ = 0.23 for this compound  -- values which are much too large and small, respectively, compared to the $J = $
4 or 9/2 values of Table 1.  If, however, we simply use the extrapolated value $\gamma =$ 0.1825 J/mol-K$^2$ we obtain $E_{max}
\gamma =$ 27.4 J/mol-K and $W=$ 1.38, values which are in much better accord with the expected scaling constants. Hence, while the
extrapolation is not very precise, it does allow us to correct the Kondo-esque scaling values in an obvious manner.\cite{Ce3In}
Note that while the antiferromagnetic contribution to the specific heat dominates $\gamma$ at low temperature, the entropy in this
contribution (defined as the excess over the extrapolated value) is a small fraction of $R ln 2$. This accords with the fact that
the spectral weight of the AF fluctuations seen in the neutron scattering is a small fraction of the weight in the primary spin
fluctuation at $E_{max}$.

 \begin{figure}[h]
\centering

\includegraphics[width=28pc]{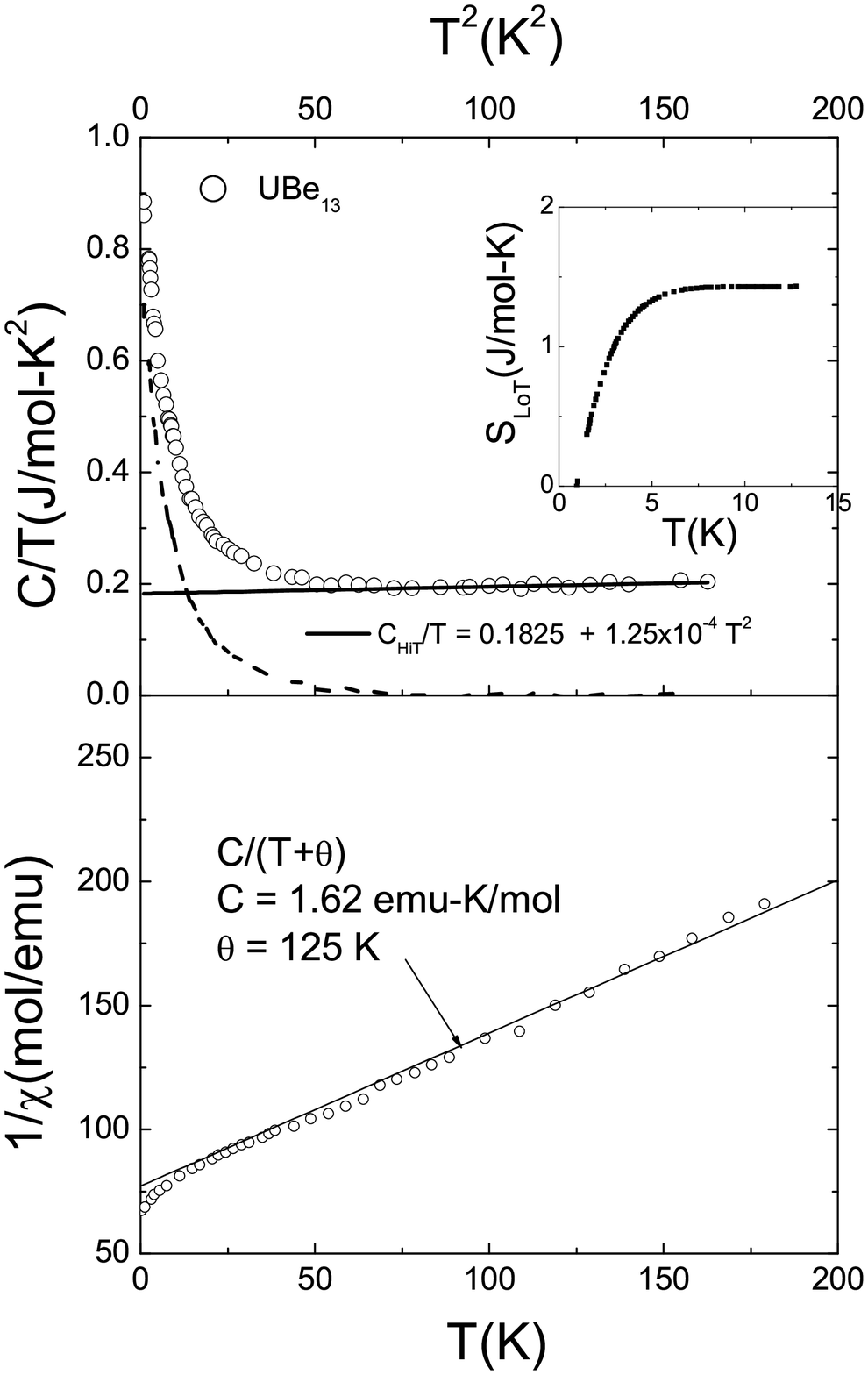}
\caption{\label{label} (a) Specific heat coefficient versus the square of the temperature for UBe$_{13}$. The solid line
extrapolates from higher temperature; the dashed line is the excess over the extrapolation, due to AF correlations. The entropy in
the excess is shown in the inset. (b) The inverse susceptibility; the solid line represents Curie Weiss behavior. (Data from Ref.
\cite{OttFisk})}

\end{figure}

 Several other uranium compounds show deviations from the scaling laws. In UPt$_3$ and UAl$_2$, the Wilson ratio is too small
 compared to the expected value, which should be of order unity, but the scaling constant $E_{max} \gamma =$ 26-34 is of the right
 order of magnitude. When the low temperature tail of the specific heat coefficient is subtracted and the extrapolated value of
 specific heat coefficient (e.g. as in Fig. 4 for UPt$_3$) is used in the scaling laws, then the Wilson ratio is corrected, but the
 $E_{max} \gamma $ is then too small. We note also that no low temperature tail of the specific heat is observed in USn$_3$, and it
 is also true for that compound that the Wilson ratio is appropriate but $E_{max} \gamma $ is too small.

 \begin{figure}[h]
\centering

\includegraphics[width=28pc]{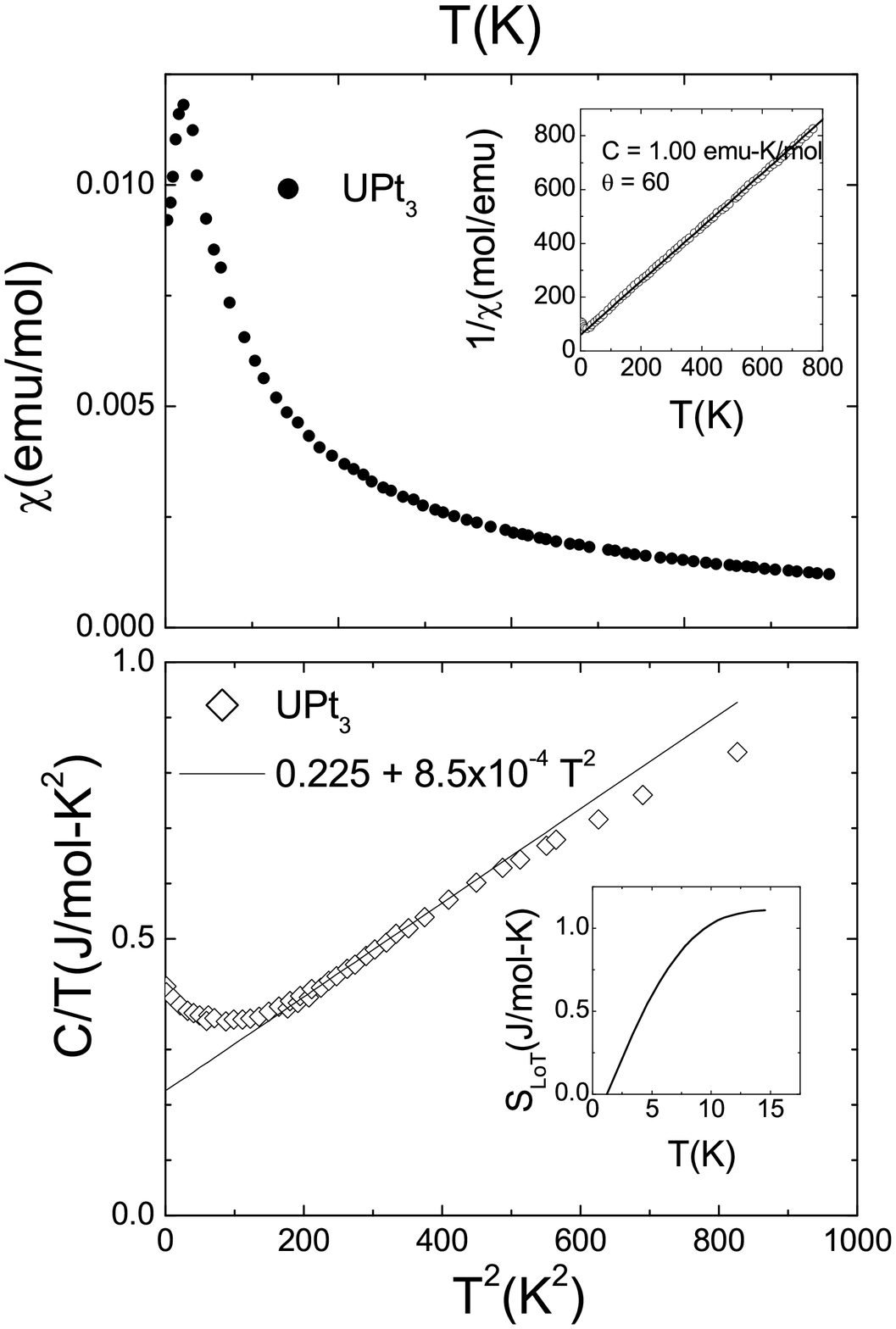}
\caption{\label{label} (a) Susceptibility of UPt$_3$; the inverse susceptibility is shown, together with a solid line representing
Curie-Weiss behavior in the inset. (b) Specific heat coefficient; the solid line is the extrapolation from high temperature. The
entropy in the excess  above the extrapolation is shown in the inset. (Data from Ref. \cite{OttFisk})}

\end{figure}

We believe that the origin of this discrepancy for these three compounds arises because the degeneracy of the 5\emph{f} moment, is
usually different from the value 9 (or 10) expected for the Hund's rule coupled $5f^2$ (or $5f^3$) configuration. This can be seen
from the fact that the high temperature Curie constant of these compounds (see the inset of Fig. 4a for UPt$_3$) is smaller than
the free ion value 1.6 emu-K/mol expected for the $5f^2$ (or $5f^3$) Hund's rule configurations. To make this statement more
graphic, we plot $T \chi (T)$ in Fig. 5 for a series of uranium compounds. This quantity appears to rapidly approach the Hund's
rule limit as temperature is increased for those compounds where the scaling either works well  (UCo$_2$Zn$_{20}$ and
URu$_2$Zn$_{20}$) or can be made to work well by correcting for the contribution from the AF fluctuations (UBe$_{13}$). For the
other compounds, it appears to saturate at a smaller value. This suppression of the high temperature moment primarily reflects the
suppression of orbital angular momentum, which is to be expected for itinerant electrons. Comparison of the 5$f$ form factor
between local moment oxides, antimonides, etc. and metallic magnets such as UNi$_2$ and UFe$_2$ (Fig. 5 inset) shows that the
orbital angular momentum is strongly reduced in the ordered state of the latter compounds.\cite{Lander} Most heavy Fermion uranium
compounds are in a correlated regime intermediate between Hund's rule local moments and purely itinerant uncorrelated $f$ electron
bands; hence we do not know the correct degeneracy to put into the scaling law for $E_{max} \gamma $. (The Wilson ratio is much
less sensitive to this degeneracy.) The degeneracy is certainly smaller than $2J+1$ for $J=$ 4 or 9/2. This probably explains why
the value  13 (17) J/mol-K obtained for this scaling constant for UPt$_3$ (UAl$_2$) after correction for the AF fluctuations (as
well as the value 10 seen for USn$_3$ without such a correction) is so  much smaller than expected for the Hund's rule $5f^2$ (or
$5f^3$) configurations.

\begin{figure}[h]
\centering

\includegraphics[width=28pc]{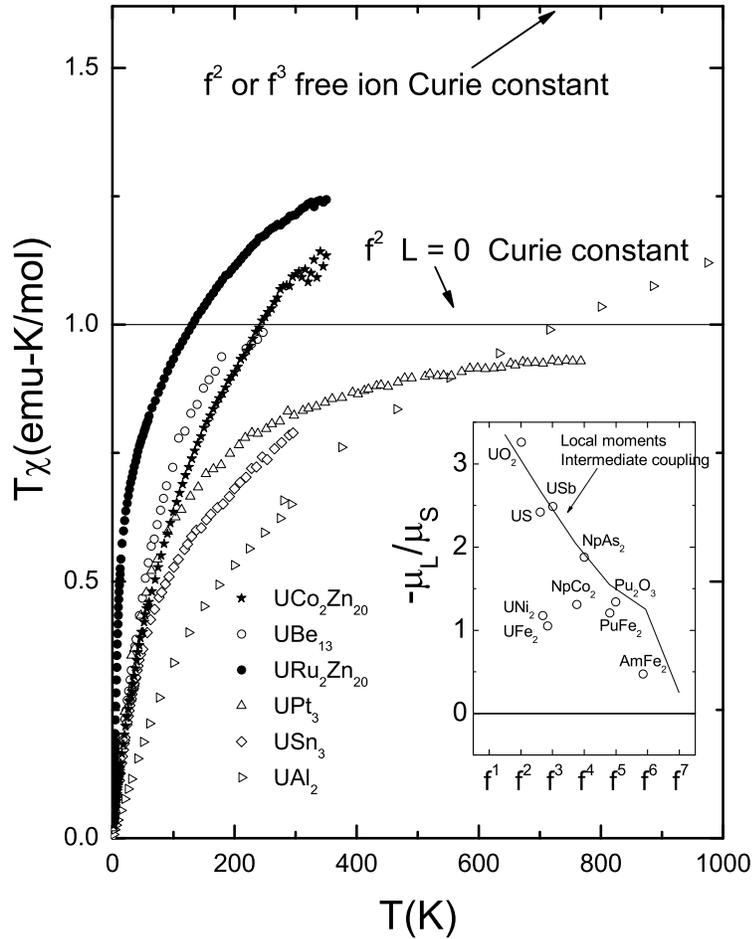}
\caption{\label{label} The quantity $T \chi (T)$ plotted versus temperature for the uranium compounds shown in Table 2. The value
of the $f^2 (J =$ 4) and $f^3 (J =$ 9/2) Curie constant is 1.62 emu-K/mol. (Sources for data given in Table 2.) Inset: The ratio of
the orbital moment to the spin moment, as determined from measurements of the 5$f$ form factor, for several uranium compounds.
Non-metallic compounds yield values of $\mu _L/\mu _S$ close to the local moment value in intermediate coupling. The intermetallic
magnets UNi$_2$ and UFe$_2$ show strong suppression of the orbital moment. (Data from Ref. \cite{Lander})}

\end{figure}

This suppression of the degeneracy of the moment associated with the spin fluctuation at $k_B T_{sf}$ is the primary way that the
itineracy of the 5$f$ electrons is reflected in the scaling laws. There are also several differences between the behavior of local
moment rare earth-based and itinerant uranium-based heavy Fermion compounds that can be deduced from this analysis. First, we note
that the energy scale ($\sim $ 100-300 K) for the spin fluctuations of the uranium heavy Fermion compounds exhibited in Table 2 is
comparable to that of cerium or ytterbium intermediate valence (IV) compounds. However, the AF fluctuations that are seen for most
of these uranium compounds are \emph{never} observed in the rare earth IV compounds, but only in very heavy Fermion compounds with small
$T_{sf} \sim $ 10 K. It is also true that there is no correlation between the magnitude of the AF contribution and $T_{sf}$ for the
uranium compounds. For example, UBe$_{13}$, which has a fairly large characteristic energy (150K), shows a large AF contribution to
the specific heat, while UPt$_3$, where $T_{sf}$ is much smaller (60K), exhibits a smaller AF contribution to $C(T)$ and USn$_3$,
for which $T_{sf} \sim$ 60K, shows no obvious upturn or AF contribution to $C(T)$.\cite{NormanUSn3} A possible explanation of these
discrepancies between rare earth and uranium compounds is that the itineracy of the 5\emph{f} electrons gives rise to an intersite
exchange that is larger relative to the spin fluctuation energy in the 5\emph{f} compounds than in the 4\emph{f} HF compounds, and
which bears no universal relation to $T_{sf}$.

From Table 2, it will be noted that the low temperature specific heat, susceptibility, and neutron characteristic energy (and hence
the low temperature scaling) are very similar for YbFe$_2$Zn$_{20}$ and UCo$_2$Zn$_{20}$. We have recently\cite{URu2Zn20} compared
the data\cite{CanfieldPNAS,BauerCo} for these two compounds to the predictions of the K/AIM and shown that the theory works
extremely well for the rare earth compound, but very poorly for the uranium compound. This disagreement undoubtedly also reflects
the difference between local moment and itinerant behavior.

Finally we point to the well-known fact that, unlike the rare earth case, well-defined crystal field excitations are almost never
observed in uranium intermetallic compounds. What $is$ seen is what we have reported above: a broad excitation on the 5-25 meV
scale, representing the primary spin fluctuation, and in some compounds low energy ($\sim$ 1 meV) antiferromagnetic fluctuations.
We thus argue that the peak at $E_{max}$ seen in uranium compounds represents spin fluctuations in an itinerant 5$f$ band.

In conclusion, we have proposed an approach that, while not highly precise, allows for separation  of the contribution of the AF fluctuations to the specific heat, thereby giving reasonable values for the Kondo-esque scaling constants. The low values (10-17) of the scaling constant $E_{max} \gamma$ observed in some uranium HF compounds arise from the low degeneracy of the high temperature moment, which is also reflected in values of the Curie constant which are small compared to the Hund's Rule values. This low degeneracy is a consequence of the itineracy of the 5$f$ electrons. 

The results allow us to formulate a line for future research: First, the quantities $C_p$, $\chi$, and both $\chi '' _{sf}$ and $\chi '' _{AF}$ (the primary and antiferromagnetic spin  fluctuations) should be measured in a single crystal of a given compound. The spectral weight of these two contributions to the magnetic neutron scattering should then be correlated with the fractional entropy of each contribution to the specific heat.
Second, the absolute cross section of the primary spin fluctuation should be measured and the appropriate sum rule used to see whether the integrated scattering corresponds to the high temperature moment seen in the susceptibility. Third,
the primary spin fluctuation needs to be measured on single crystals of more uranium compounds in an effort to clarify  the Q-dependence of this excitation in the coherent state.

\ack

Research at UC Irvine was supported by the U.S. Department of Energy, Office of Basic Energy Sciences, Division of Materials
Sciences and Engineering under Award DE-FG02-03ER46036. Work at Los Alamos National Laboratory was performed under the auspices of
the U.S. DOE/Office of Science. Work at Oak Ridge National Laboratory was sponsored by the Laboratory Directed Research and
Development Program.


\begin{thebibliography}{10}
\bibitem{CFnote} In rare earth heavy Fermions, crystal field excitations are also observed; these are not observed in rare earth IV
    compounds, nor are they observed in the heavy Fermion uranium compounds.
\end{thebibliography}
\end{document}